\documentclass[a4paper,showpacs,prb,twocolumn]{revtex4}%
\usepackage{amsfonts}
\usepackage{graphicx}
\usepackage{color}
\usepackage{amsmath}
\usepackage{amssymb}
\usepackage{latexsym}
\usepackage{psfrag}%
\begin{document}
\title{Spin polarization of edge states and magnetosubband structure in quantum wires}
\author{S. Ihnatsenka}
\altaffiliation{Permanent address: Centre of Nanoelectronics,
Department of Microelectronics, Belarusian State University for
Informatics and Radioelectronics, 220013 Minsk, Belarus}
\affiliation{Department of Science and Technology (ITN), Division
of Physics, Link\"{o}ping University, 601\thinspace74
Norrk\"{o}ping, Sweden}
\author{I. V. Zozoulenko}
\affiliation{Department of Science and Technology (ITN), Division of Physics, Link\"{o}ping
University, 601\thinspace74 Norrk\"{o}ping, Sweden}

\date{\today}

\begin{abstract}

We provide a quantitative description of the structure of edge
states in split-gate quantum wires in the integer quantum Hall
regime.  We develop an effective numerical approach based on the
Green's function technique for the self-consistent solution of
Schr\"{o}dinger equation where electron- and spin interactions are
included within the density functional theory in the local spin
density approximation. The major advantage of this technique is
that it can be directly incorporated into magnetotransport
calculations, because it provides the self-consistent eigenstates
and wave vectors at a given energy, not at a given wavevector (as
conventional methods do). We use the developed method to calculate
the subband structure and propagating states in the quantum wires
in perpendicular magnetic field starting with a geometrical layout
of the wire. We discuss how the spin-resolved subband structure,
the current densities, the confining potentials, as well as the
spin polarization of the electron and current densities evolve
when an applied magnetic field varies. We demonstrate that the
exchange and correlation interactions dramatically affect the
magnetosubbands in quantum wires bringing qualitatively new
features in comparison to a widely used model of spinless
electrons in Hartree approximation.

\end{abstract}

\pacs{73.21.Hb, 73.43.-f, 73.23.Ad}
 \maketitle

\section{Inroduction}

Transport properties of quantum dots, antidots and related structures are
affected by the nature of current-carrying states in the leads connecting
these structures to electron reservoirs. In sufficiently high magnetic fields
the current-carrying states are the edge states propagating in a close
vicinity to the sample boundaries\cite{Halperin}. A detailed information on
the structure of the edge states represent a key to the understanding of
various features of the magnetotransport in the quantum Hall regime.

A quantitative description of the edge states for the case of the gate-induced
confinement of the high-mobility two-dimensional electron gas (2DEG) was given
by Chklovskii \textit{et al. }\cite{Chklovskii}, who provided an analytical
solution for the positions and widths of the compressible and incompressible
strips arising in the 2DEG due to the electrostatic screening. In the
compressible regions, the Landau bands are pinned at the Fermi energy $E_{F}$.
This leads to a metallic behavior when the electron density is redistributed
(compressed) to keep the electrostatic potential constant. In the
incompressible regions, where the Fermi energy lies in the Landau gaps, all
the levels below $E_{F}$ are completely filled and hence the electron density
is constant (which is consistent with the behavior of the incompressible liquid).

A number of studies addressing the problem of electron-electron interaction in
quantum wires beyond the electrostatic treatment of the edge states of
Chklovskii \textit{et al. }\cite{Chklovskii} have been reported during the
recent
decade\cite{Gerhardts_1994,Kinaret,Dempsey,Tokura,Takis,Stoof,Ferconi_1995,Ando_1993,Tejedor,Ando_1998,Gerhardts_2004,Schmerek}%
. The many-body aspects of the problem have been included within
Thomas-Fermi\cite{Gerhardts_1994}, Hartree-Fock\cite{Kinaret,Dempsey,Tokura},
screened Hartree-Fock\cite{Takis}, and the density functional theory
\cite{Stoof,Ferconi_1995}. The full quantum-mechanical calculations based on
the self-consistent solution of the Schr\"{o}dinger equation have been done
within the Hartree\cite{Ando_1993,Tejedor,Ando_1998,Gerhardts_2004} and the
density functional theory\cite{Schmerek} approximations.

A particular attention has been devoted to investigation of the spin
polarization effects in edge states \cite{Kinaret,Dempsey,Tokura,Takis,Stoof}.
For example, Dempsey \textit{et al.}\cite{Dempsey}\textit{ }have shown that
for a sufficiently smooth confining potential, spin degeneracy of the
outermost edge state is lifted and two spin channels become spatially
separated. The interest to the spin-related effects in quantum wires is also
motivated by significant current activity in semiconductor spintronics, that
utilizes the spin degree of freedom of an electron to add the additional
functionality to electronic devices. A number of proposed and investigates
devices for spintronics applications operates in the edge state
regime\cite{Andy,adot}, which obviously requires a detailed knowledge of the
spatial dependence of the spin-resolved states in the quantum wires. Edge
states have also been proposed as one-way channels for transporting quantum
information\cite{Stace}. The knowledge of the spin/charge structure of the
current carrying states is also essential for numerical simulation and
modelling of the magnetotransport in quantum dots and related structures (Note
that such the modelling is often done utilizing single-electron wave functions
in the leads disregarding the spin/many electron effects\cite{Z,PRL,Europhys}%
). In order to obtain such the information on quantum-mechanical propagating
states in quantum wires, one has to solve the Sch\"{o}dinger equation
incorporating the exchange interaction to account for the spin effects. In
should be noted that the studies reported so far are often limited to some
strictly integer filling factors\cite{Dempsey,Takis}, or utilize
Thomas-Fermi-type approaches\cite{Stoof} or perturbative
technique\cite{Kinaret,Tokura}, where the required information concerning the
quantum-mechanical wave functions is not available. Moreover, the
quantum-mechanical effects associated with the finite extension of the wave
function (not included in e.g. Thomas-Fermi approach) can play a decisive role
for the quantitative description of the edge states. For example, Suzuki and
Ando\cite{Ando_1998} have demonstrated (in a model of spinless electron) that
the predictions of Chklovskii \textit{et al.} and Thomas-Fermi models
regarding the existence and the size of the compressible/incompressible strips
are in qualitative disagreement with the self-consistent modelling based on
the Schr\"{o}dinger equation in the regime when the estimated width of the
strips is smaller than the extend of the wave functions.

The purpose of the present article is two-fold. First, we perform
a detailed self-consistent solution of the Schr\"{o}dinger
equation incorporating spin/many-body effects in quantum wires. We
discuss how the spin-resolved subband structure, the current
densities, the confining potentials, as well as the spin
polarization of the electron and current densities evolve when an
applied magnetic field varies. We demonstrate that the exchange
and correlation interactions dramatically affects the
magnetosubbands in quantum wires bringing qualitatively new
features in comparison to a widely used model of spinless
electrons in Hartree approximation. In the present study we limit
ourself to the regime when more than one spin-resolved state can
propagate in the wire, i.e. the filling factor $\nu>1$ (The
filling factor $\nu =n/n_{B}=2\pi l_{B}^{2}n,$ where $n$ is the
sheet electron density, $n_{B}=eB/h$ is the number of states in
each Landau level per unit area, and
$l_{B}=\sqrt{\frac{\hbar}{eB}}$ is the magnetic length).

Second, we present a detailed description of the developed method
based on the Green's function technique for the calculation of the
subband structure and propagating states in the quantum wires in
the magnetic field. This method is numerically stable, and its
efficiency is related to the fact that calculations of the wave
functions and wave vectors are reduced to the solution of the
eigenvalue problem (as opposed to the conventional methods that
require less efficient procedure of the root
searching\cite{Ando_1993,Tejedor,Schmerek}). The major advantage
of the present method is that it can be directly incorporated into
magnetotransport calculations, because it provides the eigenstates
and wave vectors at the given energy, not at a given wavevector
(as the conventional methods do). Besides, the present method
calculates the Green's function of the wire, which can be
subsequently used in the recursive Green's function
technique\cite{Ferry,Z} widely utilized for magnetotransport
calculations in lateral structures.

In order to incorporate the spin/many-body effects into the Scr\"{o}dinger
equation we use the density functional theory (DFT) in the local spin-density
approximation\cite{ParrYang}. The choice of the DFT is motivated, on one hand,
by its efficiency and simplicity in the practical implementation within usual
self-consistent formulation introduced by Kohn and Sham\cite{Kohn_Sham}, and,
on the other hand, by its success in the reproduction of the electronic and
spin properties of the low-dimensional structures in comparison to the exact
diagonalization and quantum Monte-Carlo calculations, as well as experiments
(for a review, see \cite{QDOverview}). For example, Ferconi and
Vignale\cite{Ferconi_Vignale} find that the accuracy of the DFT for the energy
and density of few-electron quantum dots yields the accuracy better than 3\%
in comparison to the exact results. An excellent agreement between DFT and the
variational Monte-Carlo results for the chemical potential and the addition
spectra of the rectangular quantum dot was reported by R\"{a}s\"{a}nen
\textit{et al.}\cite{Rasanen}.

Within the local spin density approximation the exchange and correlation
potentials are calculated using a parameterization of the functional for the
exchange and correlation energy $\epsilon_{xc}.$ The latter is usually
obtained on the basis of quantum Monte Carlo calculations
\cite{TC,Attaccalite} for corresponding infinite homogeneous system. In the
present paper we use the parameterization of Tanatar and Cerperly (TC)
\cite{TC}. This parameterization is valid for magnetic field when $\nu>1$,
which defines the range of applicability of our results. (Various
parameterizations for $\epsilon_{xc}$ for strong fields $\nu<1$ as well as
different interpolation schemes between the low and the strong fields are
reviewed in Refs. \cite{QDOverview,Saarikoski}). Note that the DFT was used
for the description of the spin polarization of the edge states in quantum
wires in the integer Hall regime within Thomas-Fermi approximation\cite{Stoof}%
, as well as for the treatment of spinless edge states in the Kohn-Sham scheme
based on the solution of the Schr\"{o}dinger equation\cite{Schmerek}. The
density functional theory within the Thomas-Fermy approach was also applied
for the description of the edge channels in the quantum wire in the fractional
Hall regime, where the parameterization of $\epsilon_{xc}$ incorporated the
additional gaps that open up at the fractional filling
factors\cite{Ferconi_1995}.

The paper is organized as follows. In Section II we present a
formulation of the problem, where we define the geometry of the
system at hand and outline the self-consistent Kohn-Sham scheme
within the LSDA approximation. In Section III we provide a
detailed description of or method based on the Green's function
technique, and Section IV presents the major results and their
discussion. The conclusions are given in Section V, and Appendix
presents some technical details of the calculations.

\section{Formulation of the problem}

\begin{figure}[ptb]
\includegraphics[scale=0.6]{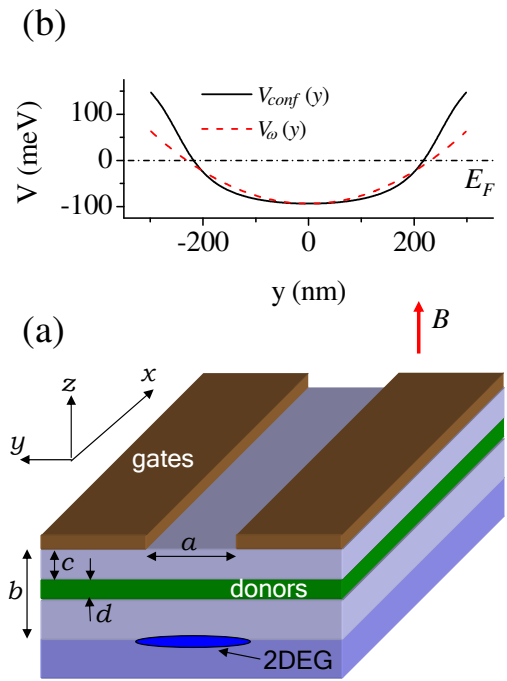}
\caption{(Color online). (a) A schematic layout of a split-gate
quantum wire in a perpendicular magnetic field. (b) Solid line:
the calculated electrostatic potential $V_{conf}(y)$ for the
quantum wire with $a=500$ nm, $b=60$ nm, $c=14$ nm, $d=$ 36 nm,
$\rho_{d}=6\cdot10^{23}$ m$^{-3},$ and $|V_{0}|=0.2$ V. A dashed
line shows the corresponding parabolic confinement
$V_{\omega}(y)=m^{\ast}\left(\omega y\right)^{2}/2$ (with
$\hbar\omega=2meV)$ often used
to approximate the external electrostatic confinement.}%
\label{fig:wire}%
\end{figure}

We consider an infinitely long split-gate quantum wire in a perpendicular
magnetic field. A schematic layout of the device is illustrated in Fig.
\ref{fig:wire} (a).
The distance between gates is $a,$ the distance from the surface to the
electron gas is $b$ (we disregard the spatial extension of the electron wave
function in the $z$-direction). The donor layer with the donor density
$\rho_{d}$ has the width $d$ and is situated at the distance $c$ from the
surface. The external electrostatic confinement potential can be written in
the form%
\begin{equation}
V_{conf}(y)=V_{g}(y)+V_{d}+V_{Schottky}, \label{el}%
\end{equation}
where $V_{g}(y)$ and $V_{d}$ are respectively potentials due to the
gates\cite{Davies} and the donor layer\cite{Martorell}, and $V_{Scho}$ is the
Schottky barrier,%
\begin{align}
V_{g}  &  =|V_{0}|\left\{  1-\frac{1}{\pi}\left[  \arctan\frac{a+y}{b}%
+\arctan\frac{a-y}{b}\right]  \right\}  ,\label{Vg}\\
V_{d}  &  =-\frac{e^{2}}{\varepsilon_{0}\varepsilon_{r}}\rho_{d}d\left(
c+d/2\right)  , \label{Vd}%
\end{align}
with $V_{0}$ being the (negative) applied gate voltage, and $\varepsilon_{r}$
being the dielectric constant. The Schottky potential is chosen to be
$V_{Schottky}=0.8$ eV, which is appropriate for GaAs. The external
electrostatic confinement potential is shown in Fig. \ref{fig:wire} (b) for a
representative quantum wire with parameters typical for an experiment. Figure
\ref{fig:wire} (b) also illustrates the corresponding parabolic potential
$V(y)=\frac{m^{\ast}}{2}\left(  \omega y\right)  ^{2}$ often used to
approximate the electrostatic confinement in the split-gate wires, where
$m^{\ast}$ is the effective electron mass ($m^{\ast}=0.067m_{e}$ for GaAs).

The wire is described by the effective Hamiltonian in a perpendicular magnetic
field, $\mathbf{B_{\bot}}=B\mathbf{\hat{z}}$,
\begin{equation}
H^{\sigma}=H_{0}+V_{conf}(y)+V_{eff}^{\sigma}(y)+g\mu_{b}B\sigma,
\label{Hamiltonian}%
\end{equation}
where $H_{0}$ is the kinetic energy in the Landau gauge, $\mathbf{A}%
=(-By,0,0)$,
\begin{equation}
H_{0}=-\frac{\hbar^{2}}{2m^{\ast}}\left\{  \left(  \frac{\partial}{\partial
x}-\frac{eiBy}{\hbar}\right)  ^{2}+\frac{\partial^{2}}{\partial^{2}y}\right\}
. \label{H_0}%
\end{equation}
The last term in Eq. (\ref{Hamiltonian}) accounts for Zeeman energy where
$\mu_{b}=e\hbar/2m_{e}$ is the Bohr magneton, $\sigma=\pm%
\frac12
$ describes spin-up and spin-down states, $\uparrow$ , $\downarrow$, and the
bulk $g$ factor of GaAs is $g=-0.44.$

The effective potential, $V_{eff}^{\sigma}(y)$ within the framework of the
Kohn-Sham density functional theory reads \cite{ParrYang,Kohn_Sham,QDOverview}%
,
\begin{equation}
V_{eff}^{\sigma}(\mathbf{r})=V_{H}(y)+V_{xc}^{\sigma}(y). \label{V_eff}%
\end{equation}
$V_{H}($\textbf{r}$)$ is the Hartree potential due to the electron density
$n(y)=\sum_{\sigma}n^{\sigma}(y)$ (including the mirror charges),%
\begin{align}
&  V_{H}(y)=\frac{e^{2}}{4\pi\varepsilon_{0}\varepsilon_{r}}\int_{-\infty
}^{+\infty}dx^{\prime}\int_{-\infty}^{+\infty}dy^{\prime}n(y)\nonumber\\
&  \ \times\Bigg[\frac{1}{\sqrt{\left(  x-x^{\prime}\right)  ^{2}+\left(
y-y^{\prime}\right)  ^{2}}}-\nonumber\\
&  -\frac{1}{\sqrt{\left(  x-x^{\prime}\right)  ^{2}+\left(  y-y^{\prime
}\right)  ^{2}+4b^{2}}}\Bigg]\nonumber\\
&  =-\frac{e^{2}}{4\pi\varepsilon_{0}\varepsilon_{r}}\int_{-\infty}^{+\infty
}dy^{\prime}n(y^{\prime})\ln\frac{\left(  y-y^{\prime}\right)  ^{2}}{\left(
y-y^{\prime}\right)  ^{2}+4b^{2}}. \label{V_H}%
\end{align}
The exchange and correlation potential $V_{xc}(y)=V_{x}(y)+V_{c}(y)$ in the
local spin density approximation is given by%
\begin{equation}
V_{xc}^{\sigma}=\frac{d}{dn^{\sigma}}\left\{  n^{\sigma}\epsilon_{xc}\left(
n,\zeta(y)\right)  \right\}  \label{LDA}%
\end{equation}
where
$\zeta(y)=\frac{n^{\uparrow}-n^{\downarrow}}{n^{\uparrow}+n^{\downarrow
}}$ is the local spin-polarization. As we mentioned in
Introduction, in the present paper we use the parameterization of
Tanatar and Cerperly (TC) \cite{TC}; for the sake of completeness,
the explicit expressions for $V_{x}(y)$ and $V_{c}(y)$ are given
in Appendix (see also Ref. [\onlinecite{Macucci1993}]).

\section{Calculation of the electron density and edge states in quantum
wires.}

In order to calculate the self-consistent electron densities, wave functions
and wave vectors of the magneto-edge states as well as corresponding currents,
we use the Green's function technique. A detailed account of the major steps
of the calculations is presented in this section.

\subsection{\textit{Hamiltonian in the mixed energy-space representation.}}

Numerical computation of the self-consistent electron densities and other
quantities of interest requires the discretization of the Hamiltonian
(\ref{Hamiltonian}). Introduce a numerical grid (lattice) with the discrete
variables $m,n$ according to $x,y\rightarrow ma,na,$ where $a$ is the lattice
constant. The computational domain consists of $N_{s}$ sites in the transverse
$n$-direction (the wire is infinite in the longitudinal $m$-direction).
Discretization of the continuous Hamiltonian (\ref{Hamiltonian}) gives a
standard tight-binding Hamiltonian with the magnetic field included in the
form of the Peierls substitution\cite{Ferry},%
\begin{align}
H^{\sigma} &  =\sum_{m}\Bigg\{\sum_{n=1}^{N_{s}}\left\{  \epsilon
_{0}+V^{\sigma}(n)\right\}  a_{m,n}^{+}a_{m,n}-\label{tight-binding}\\
&  -t\,\left\{  a_{m,n}^{+}a_{m,n+1}+e^{-iqn}a_{m,n}^{+}a_{m+1,n}%
+\text{h.c.}\right\}  \Bigg\},\nonumber
\end{align}
where
\begin{equation}
V^{\sigma}(n)=V_{conf}(n)+V_{eff}^{\sigma}(n)+g\mu_{b}B\sigma\label{V_total}%
\end{equation}
is the total confining potential, the hopping element $t=\hbar^{2}/2m^{\ast
}a^{2},$ the site energy $\epsilon_{0}=4t,$ and $q=eBa^{2}/\hbar;$
$a_{m,n}^{+}$ and $a_{m,n}$ denote the creation and annihilation operators at
the site $\left(  m,n\right)  .$ The translational invariance in the
longitudinal direction dictates the Bloch form for the propagating states in
the quantum wire,%
\begin{equation}
\left\vert \psi_{\alpha}^{\sigma}\right\rangle =\sum_{m}e^{ik_{\alpha}%
^{\sigma}m}\sum_{n=1}^{N_{s}}\psi_{\alpha}^{\sigma}(n)\,a_{m,n}^{+}\left\vert
0\right\rangle ,\label{Bloch_real_space}%
\end{equation}
where the index $\alpha$ corresponds to the $\alpha$-th Bloch state with the
wave vector $k_{\alpha}^{\sigma}$ and the transverse wave function
$\psi_{\alpha}^{\sigma}(n).$ In Eq. (\ref{Bloch_real_space}) the wave function
$\psi_{\alpha}^{\sigma}(n)$ corresponds to the real space representation. To
facilitate the numerical calculation, it is convenient to expand the
wavefunctions over the transverse eigenstates (modes) of a homogeneous wire of
the width of $N_{s}$ sites, $\phi_{j}(n)=\sqrt{\frac{2}{N_{s}+1}}\sin\frac{\pi
jn}{N_{s}+1},$%
\begin{equation}
\psi_{\alpha}^{\sigma}(n)=\sum_{j=1}^{N}\psi_{\alpha,j}^{\sigma}\phi
_{j}(n),\label{expansion}%
\end{equation}
where the expansion coefficients $\psi_{\alpha,j}^{\sigma}$ can be interpreted
as the wavefunction in the \textquotedblleft energy\textquotedblright%
\ representation in the space of the transverse eigenstates\cite{Z}. Note that
Eq. (\ref{expansion}) corresponds to a conventional sin-transformation, whose
inverse transform is given by the same equation\cite{Num_rec}. The summation
in Eq. (\ref{expansion}) runs over $1\leq j\leq N$, with $N=N_{s}.$ In
practice, however, it is sufficient to limit the summation to much smaller
number of modes, with $N\ll N_{s}.$ Because the speed of the method is
determined by the dimension of the matrices (that is given by $N_{s}$ in the
real space representation and $N$ in the \textquotedblleft
energy\textquotedblright\ representation), passing to the \textquotedblleft
energy\textquotedblright\ representation greatly enhances the computational
speed. For example, for the wire of the width of 0.5 $\mu$m, it is sufficient
to use $N\approx50$ modes to achieve a good convergence of the results with
respect to the mode number. At the same time, in the real space
representation, $N_{s}=500$ (with the lattice constant $a=1$ nm), which makes
computations rather impractical.

Passing from the real space representation to the \textquotedblleft
energy\textquotedblright\ representation in the transverse direction we arrive
to the Hamiltonian in the mixed energy-space representation (i.e., in the real
space representation in the longitudinal $m$-direction and \textquotedblleft
energy\textquotedblright\ representation in the transverse $n$%
-direction)\cite{Z},
\begin{align}
H  &  =\sum_{m}\Bigg\{\sum_{j=1}^{N}\left\{  \epsilon_{j}+2t\right\}
a_{m,j}^{+}a_{m,j}+\sum_{j,j^{\prime}}^{N}V_{jj^{\prime}}^{\sigma}a_{m,j}%
^{+}a_{m,j^{\prime}}\nonumber\\
&  -\sum_{j,j^{\prime}}^{N}\left[  t_{jj^{\prime}}^{L}a_{m,j}^{+}%
a_{m+1,j^{\prime}}+t_{jj^{\prime}}^{R}a_{m+1,j}^{+}a_{m,j^{\prime}}\right]
\Bigg\}, \label{H_energy_representation}%
\end{align}
where $\epsilon_{j}=2t-2t\cos\frac{\pi j}{N+1}$ are the eigenvalues of the
transverse motion corresponding to the eigenfunctions $\phi_{j}(n);$ the
creation and annihilation operators in the mixed space-energy representation
are related to the real space creation and annihilation operators according to
$a_{m,j}^{+}=\sum_{n=1}^{N_{s}}\phi_{j}(n)a_{m,n}^{+},$ $a_{m,j}=\sum
_{n=1}^{N_{s}}\phi_{j}(n)a_{m,n}$. The matrix elements of total confining
potential and the hopping matrix elements read are given by%
\begin{align}
V_{jj^{\prime}}^{\sigma}  &  =\sum_{n=1}^{N_{s}}\phi_{j}(n)V^{\sigma}%
(n)\phi_{j^{\prime}}(n),\;\label{V_U}\\
t_{jj^{\prime}}^{R}  &  =t\sum_{n=1}^{N_{s}}\phi_{j}(n)e^{iqn}\phi_{j^{\prime
}}(n),\;t_{jj^{\prime}}^{L}=(t_{jj^{\prime}}^{R})^{\ast}.\nonumber
\end{align}
Note that Hamiltonian (\ref{H_energy_representation}) has nearest-neighbor
couplings in the longitudinal $m$-direction (described by two last terms in
Eq. (\ref{H_energy_representation})). In the transverse (\textquotedblleft
energy\textquotedblright) direction, the magnetic field couples all states $j$
on slice $m$ to all states $j\prime$ on neighboring slices $m+1$ and $m-1$.
The Bloch wavefunctions (\ref{Bloch_real_space}) in the mixed space-energy
representation read%
\begin{equation}
\left\vert \psi_{\alpha}^{\sigma}\right\rangle =\sum_{m}e^{ik_{\alpha}%
^{\sigma}m}\sum_{j=1}^{N}\psi_{\alpha,j}^{\sigma}\,a_{m,j}^{+}\left\vert
0\right\rangle . \label{Bloch_mixed_representation}%
\end{equation}

\subsection{\textit{Bloch states of a quantum wire in magnetic field.}}

\begin{figure}[ptb]
\includegraphics[scale=0.6]{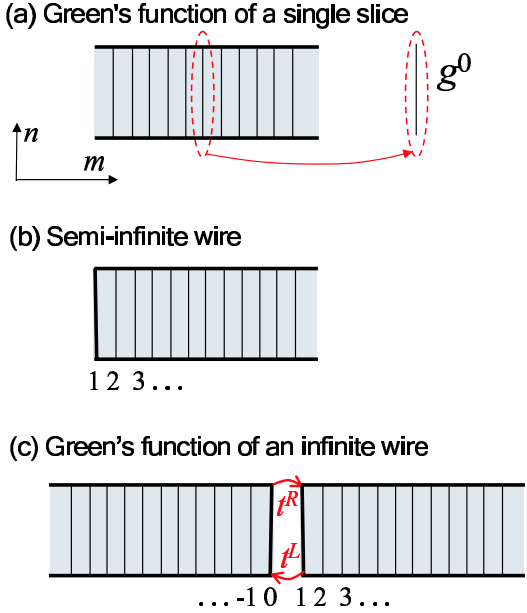}
\caption{(Color online). Graphical illustration of the calculations of the
Green's function for a single slice (a), and for an infinite wire (b), (c).}%
\label{fig:graphics}%
\end{figure}

Define a retarded Green's function of the Hamiltonian $H$ in a standard
way\cite{Ferry,Datta},%
\begin{equation}
\left(  E-H+i\varepsilon\right)  G=\mathbf{1,} \label{Green's_function}%
\end{equation}
where $\mathbf{1}$ is a unitary operator. Calculate first the Green's function
$g^{\sigma}$ corresponding to a single slice (see Fig. \ref{fig:graphics}
(a)).
The Hamiltonian of $m$-th single slice reads
\begin{equation}
h_{m}^{\sigma}=\sum_{j=1}^{N}\left\{  \epsilon_{j}+2t\right\}  a_{m,j}%
^{+}a_{m,j}+\sum_{j,j^{\prime}}^{N}V_{jj^{\prime}}^{\sigma}a_{m,j}%
^{+}a_{m,j^{\prime}}, \label{H_single_slice}%
\end{equation}
(Note that a single slice is not coupled to its neighbors, and hence two last
terms in Eq. (\ref{H_energy_representation}) are absent in Eq.
(\ref{H_single_slice})). Using this operator in the definition of Green's
function (\ref{Green's_function}), and taking the matrix elements
$\langle0|a_{m,j}...\,a_{m,j^{\prime}}^{+}|0\rangle,$ we arrive to the
$N\times N$ system of linear equations for the matrix elements of the Green's
function of a single slice $g_{j^{\prime\prime}j^{\prime}}^{\sigma}%
=\langle0|a_{m,j}\,g^{\sigma}\,a_{m,j^{\prime}}^{+}|0\rangle$,%
\begin{equation}
\sum_{j^{\prime\prime}=1}^{N}\left(  \left(  E-\epsilon_{j}-2t\right)
\delta_{j,j^{\prime\prime}}-V_{jj^{\prime\prime}}^{\sigma}\right)
g_{j^{\prime\prime}j^{\prime}}^{\sigma}=\delta_{j,j^{\prime}}.
\label{single_slice}%
\end{equation}
Note that because of the translational invariance, we have dropped index $m$
in the definition of the matrix element of the Green's function of a single slice.

Knowledge of the Green's function of a single slice $g^{\sigma}$ allows one to
find the Bloch states of an infinite wire. The eigenvectors $\left\{
\psi_{j\alpha}^{\sigma}\right\}  $ and eigenvalues $\left\{  k_{\alpha
}^{\sigma}\right\}  $ are determined by the eigenequation\cite{Z}
\begin{equation}
\left(
\begin{array}
[c]{cc}%
-\left(  g^{\sigma}t^{L}\right)  ^{-1} & -\left(  t^{L}\right)  ^{-1}t^{R}\\
1 & 0
\end{array}
\right)  \left(
\begin{array}
[c]{c}%
e^{ik}\overrightarrow{\psi^{\sigma}}\\
\overrightarrow{\psi^{\sigma}}%
\end{array}
\right)  =e^{ik}\left(
\begin{array}
[c]{c}%
e^{ik}\overrightarrow{\psi^{\sigma}}\\
\overrightarrow{\psi^{\sigma}}%
\end{array}
\right)  , \label{eigenequation}%
\end{equation}
where the matrixes $t^{R},t^{L}$ and $g^{\sigma}$ have matrix elements given
by Eqs. (\ref{V_U}) and (\ref{single_slice}) respectively, and
$\overrightarrow{\psi^{\sigma}}$ is the column vector composed of $\psi
_{j}^{\sigma},$ $1\leq j\leq N$ (Eq. (\ref{expansion})). Here and hereafter we
use Greek indexes $\alpha,$ $\alpha^{\prime}$ for Bloch states of the wire,
and Roman indexes $j,j^{\prime}$ for the basis set of the transverse
eigenfunctions $\left\{  \phi_{j}(n)\right\}  $. Equation (\ref{eigenequation}%
) has $2N$ eigenvalues $k_{\alpha}^{\sigma},$ $1\leq\alpha\leq N$, which can
be real or complex, describing respectively propagating and evanescent states
(Here and hereafter $k_{\alpha}^{\sigma}$ is given in units of $a^{-1}$ and
the group velocity $v$ is in units of $a)$. The eigenvalues corresponding to
right propagating states ($v=\,\partial E/\partial k>0,$ $\Im(k)=0$) and
states decaying to the right ($\Im(k)>0$) we denote by $k_{\alpha}^{\sigma
^{+}}$ with corresponding eigenstates $\psi_{j\alpha}^{\sigma^{+}}$.
Correspondingly, $k_{\alpha}^{\sigma^{-}}$ and $\psi_{j\alpha}^{\sigma^{-}}$
stand for left propagating states ($v=\,\partial E/\partial k<0,$ $\Im(k)=0$)
and states decaying to the left ($\Im(k)<0$). Sorting right- and
left-propagating eigenstates can be easily done by calculating their group
velocity\cite{Z}
\begin{equation}
v_{\alpha}^{\sigma}=\frac{1}{\hbar}\frac{\partial E}{\partial k_{\alpha
}^{\sigma}}=-\frac{2}{\hbar}\sum_{j,j^{\prime}}^{N}\left(  \psi_{j\alpha
}^{\sigma}\right)  ^{\ast}\psi_{j^{\prime}\alpha}^{\sigma}\Im\left[
e^{-ik_{\alpha}^{\sigma}}t_{jj^{\prime}}^{L}\right]  . \label{velocity}%
\end{equation}
Passing to the real space representation for the wave functions in the above
expressions and using the quantum-mechanical particle current density for
$\alpha$-th Bloch state calculated in a standard way for a tight-binding
lattice\cite{Z},%
\begin{equation}
j_{\alpha}^{\sigma}(n,E)=\frac{2t}{\hbar}\sin\left(  qn+k_{\alpha}^{\sigma
}\right)  |\psi_{\alpha}^{\sigma}(n)|^{2}, \label{QM_current}%
\end{equation}
the group velocity (\ref{velocity}) can be expressed as the total particle
current of $\alpha$-th Bloch state,%
\begin{equation}
v_{\alpha}^{\sigma}=\sum_{n}j_{\alpha}^{\sigma}(n,E). \label{velocity_j}%
\end{equation}
To conclude this section we note that a direct calculation of the
eigenvectors $k_{\alpha}^{\sigma}$ and eigenfunctions
$\psi_{j^{\prime}\alpha}^{\sigma}$ by substitution of Eq.
(\ref{Bloch_mixed_representation}) into Schr\"{o}dinger equation
and calculation of the roots of the corresponding determinant is
possible (see e.g., Refs.
\onlinecite{Ando_1993,Tejedor,Gerhardts_2004}). However, the
procedure used here is more efficient as the solution of the
eigenproblem (\ref{eigenequation}) is numerically faster and less
demanding than the  root searching. Besides, an important
advantage of the present method is that it can be directly
incorporated into magnetotransport calculations, because in
contrast to the root searching method, the present technique
provides the eigenstates and wave vectors at the given energy, not
at a given wavevector.

\subsection{\textit{Calculation of the local electron density.}}

The diagonal elements of the total Green's function of an infinite
wire  in the real space representation give the local density of
states (LDOS) at the site
$\mathbf{r}=(m,n)\cite{Datta},$%
\begin{equation}
\rho^{\sigma}(\mathbf{r},E)=-\frac{1}{\pi}\Im\left[  G^{\sigma}(\mathbf{r}%
,\mathbf{r},E)\right]  . \label{LDOS}%
\end{equation}

\begin{figure}[tb]
\includegraphics[scale = 0.6]{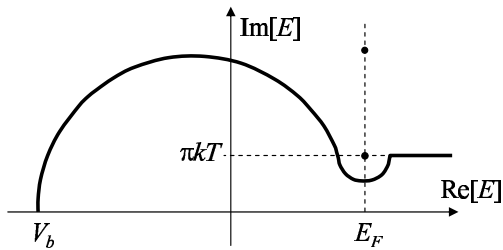}\caption{A typical integration contour
used in the calculation of integral (\ref{n}). Dots indicate the poles of the
Fermi-Dirac distribution function in the upper complex plane at $\Re\left[
E\right]  =E_{F}$, $\Im\left[  E\right]  =(2m+1)\pi kT,$ $m=0,1,2,\ldots$ .}%
\label{fig:contour}%
\end{figure}

The LDOS $\rho^{\sigma}(\mathbf{r},\mathbf{r},E)$ can be used to calculate the
local electron density at the site $\mathbf{r}$,
\begin{equation}
n^{\sigma}(\mathbf{r})=\int_{V_{b}}^{\infty}dE\,\rho^{\sigma}(\mathbf{r}%
,E)f(E-E_{F}), \label{n}%
\end{equation}
where $f(E-E_{F})$ is the Fermi-Dirac distribution function and the lower
limit of integration $V_{b}$ corresponds to the bottom of the total confining
potential. Note that $\rho^{\sigma}(\mathbf{r},\mathbf{r},E)$ is a rapidly
varying function of energy diverging as $\sim\left(  E-E_{\alpha}\right)
^{-1/2}$ when $E$ approaches the threshold subband energies $E_{\alpha}$.
Because of this, a direct integration along the real axis is rather
ineffective as its numerical accuracy is not sufficient to achieve convergence
of the self-consistent calculation of the electron density . We therefore
calculate integral (\ref{n}) by transforming the integration contour into the
complex energy plane $\Im\left[  E\right]  >0$ where the Green's function is
much more smoother than on the real axis. (Note that all poles of the Green's
function (\ref{LDOS}) are in the lower half-plain $\Im\left[  E\right]  <0$).
A typical contour used in the integration avoiding poles of the Fermi-Dirac
function $f(E-E_{F})$ is shown in Fig. \ref{fig:contour}.
We calculate the diagonal elements of the total Green's function $G^{\sigma
}(\mathbf{r},\mathbf{r},E)$ as follows. We start from a semi-infinite quantum
wire and calculate its surface Green's function $\Gamma$ (i.e. Green's
function for the boundary slice $m=1$), see Fig. \ref{fig:graphics} (b). The
right and left surface Green's functions $\Gamma^{r}$ and $\Gamma^{l}$
(corresponding to a semi-infinite wires open respectively to the right and
left) can be written in a matrix form\cite{Ando_Gamma,Z}
\begin{align}
\Gamma_{\sigma}^{r}t^{R}  &  =-\Psi_{\sigma}^{+}K_{\sigma}^{+}\Psi_{\sigma
}^{+^{-1}}\label{Gamma}\\
\Gamma_{\sigma}^{l}t^{L}  &  =-\Psi_{\sigma}^{-}\left(  K_{\sigma}^{-}\right)
^{-1}\left(  \Psi_{\sigma}^{-}\right)  ^{-1},\nonumber
\end{align}
where the matrix elements $(\Psi_{\sigma}^{+(-)})_{j\alpha}=$ $\psi_{j\alpha
}^{\sigma^{+(-)}},$ $(K_{\sigma}^{+(-)})_{\alpha\alpha^{\prime}}=\exp\left(
k_{\alpha}^{\sigma^{+(-)}}\right)  \delta_{\alpha\alpha^{\prime}}.$ We then
connect this semi-infinite wire to the second semi-infinite wire to form an
infinitely long quantum wire as shown in Fig. \ref{fig:graphics} (c). The
total Green's function can be calculated with the help of Dyson
equation\cite{Ferry,Datta}%
\begin{equation}
G=G^{0}+G^{0}UG, \label{Dyson}%
\end{equation}
where $G^{0}$ correspond to the \textquotedblleft
unperturbed\textquotedblright\ structures (the left and right semi-infinite
wires), and the operator $U$ describes the interaction between them,
\begin{equation}
U=-\sum_{j,j^{\prime}}^{N}\left[  t_{jj^{\prime}}^{L}a_{0,j}^{+}%
a_{1,j^{\prime}}+t_{jj^{\prime}}^{R}a_{1,j}^{+}a_{0,j^{\prime}}\right]  ,
\label{perturbation}%
\end{equation}
(see Eq. (\ref{H_energy_representation})). Using Eq. (\ref{Dyson}) to
calculate the matrix element $(G^{\sigma})_{jj^{\prime}}=\langle
0|a_{1,j}\,G^{\sigma}\,a_{1,j^{\prime}}^{+}|0\rangle$, we obtain Green's
function for the slice $m=1$ (Note that because of the translational
invariance in the $m$-direction, the calculated Green's function is the same
for all slices),
\begin{equation}
G^{\sigma}=\left(  \mathbf{1}-\Gamma_{\sigma}^{r}t^{R}\Gamma_{\sigma}^{l}%
t^{L}\right)  ^{-1}\Gamma_{\sigma}^{r}, \label{G_total}%
\end{equation}
where $\mathbf{1}$ is the unit matrix. Note that Eq. (\ref{G_total}) gives
Green's functions in the \textquotedblleft energy\textquotedblright%
\ representation of the space of the transverse eigenstates. To obtain the
Green's function in the real space representation needed to compute the
electron density $n^{\sigma}(\mathbf{r})$ (\ref{n}) we perform a standard
change of the basis, $G^{\sigma}(n,n,E)=\sum_{j,j^{\prime}}^{N}\varphi
_{j}(n)G_{jj^{\prime}}^{\sigma}\varphi_{j^{\prime}}(n).$

\subsection{\textit{Self-consistent calculations}}

\textit{Iteration procedure}. We calculate magneto-edge states and electron
densities in a quantum wire in a self-consistent way, when on each iteration
step a small part of a new potential (\ref{V_total}) is mixed with the old one
(from the previous iteration step),
\begin{equation}
V_{i+1}^{\sigma}(n)=(1-\epsilon)V_{i}^{\sigma}(n)+\epsilon\,V_{i+1}^{\sigma
}(n),\label{mixing_V}%
\end{equation}
$\epsilon$ being a small constant, $\sim 0.1 - 0.01$. Using this
input potential we, for a given energy $E,$ solve the eigenproblem
(\ref{eigenequation}) to find the Bloch states in the quantum
wire. (Note that energy $E$ is chosen in the complex plane as
shown in Fig. (\ref{fig:contour})). We then use the obtained
results to calculate the Green's function $G^{\sigma}(n,n,E)$
according to Eqs. (\ref{Gamma})-(\ref{G_total}). The integration
of the Green's function (\ref{n}) gives the electron densities
$n^{\sigma}(\mathbf{r})$, which are
subsequently used to compute the new total confining potential (\ref{V_total}%
). It is typically needed $\sim1000$ iteration steps to achieve our
convergence criterium  $|n_{1D}^{i+1}-n_{1D}^{i}|/(n_{1D}^{i+1}+n_{1D}%
^{i})<10^{-5},$ where $n_{1D}^{i}$ is the one dimensional electron density
$n_{1D}=\int n(\mathbf{r})\,dy\,$ on $i$-th iteration step.

\textit{Adjustment of the Fermi energy}. When the same fixed Fermi energy
$E_{F}$ is used for different magnetic fields $B$, the calculated
self-consistent one dimensional electron density changes as $B$ varies$.$
Depending on a particular realization of a quantum wire, one might need to
adjust $E_{F}$ for each $B$ in order to keep the total electron density fixed,
as magnetic field does not change the electron density in the system. However,
in a typical experimental situation when a long quantum wire is connected to a
2DEG \cite{Wrobel}, the Fermi energy in the reservoirs (not the electron
density in the wire) is fixed. Because of this, in all calculations reported
in the paper we keep $E_{F}$ fixed (we set $E_{F}=0$).

Note that we have also performed calculations where $E_{F}$ was adjusted to
keep the electron density $n_{1D}^{\sigma}$ constant. All the results obtained
in this case (in particularly, the density and current spin polarizations) are
qualitatively and quantitatively similar to those obtained in the case when
$E_{F}$ is adjusted.

\textit{Bloch states, subband structure and current density}. Having
calculated the total self-consisted confining potential, we can compute the
Bloch wave functions and wave vectors by solving the eigenequation
(\ref{eigenequation}) for the whole range of energies of interest (note that
for these calculations the energy has to be chosen on the real axis).
Knowledge of the wave vectors for different states allows us to recover the
subband structure, i.e. to calculate an overage position $x_{\alpha}^{\sigma}$
of the wave functions for different modes $\alpha$\cite{Davies_book},
\begin{equation}
x_{\alpha}^{\sigma}=\frac{\hbar k_{\alpha}^{\sigma}a}{eB}. \label{x_b}%
\end{equation}

We calculate the conductance of the wire $G^{\sigma}=I^{\sigma}/V$ on the
basis of the linear-response Landauer formula,%
\begin{equation}
I^{\sigma}=\frac{e^{2}}{h}V\sum_{\alpha}\int_{E_{th\,\alpha}^{\sigma}}%
^{\infty}dE\left(  -\frac{\partial f\left(  E-E_{F}\right)  }{\partial
E}\right)  , \label{I}%
\end{equation}
where summation is performed over all propagating modes $\alpha$ for the spin
$\sigma,$ with $E_{th\,\alpha}^{\sigma}$ being the propagation threshold for
$\alpha$-th mode. In order to visualize the current density we can re-write
Eq. (\ref{I}) for the total current in the form $I_{\alpha}^{\sigma}=a\sum
_{n}J_{\alpha}^{\sigma}(n)\,$, where the current density for the mode $\alpha$
reads%
\begin{equation}
J_{\alpha}^{\sigma}(n)=\frac{e^{2}}{h}V\int dE\frac{j_{\alpha}^{\sigma}%
(n,E)}{v_{\alpha}^{\sigma}}\left(  -\frac{\partial f\left(  E-E_{F}\right)
}{\partial E}\right)  , \label{J}%
\end{equation}
with $v_{\alpha}^{\sigma}$ and $j_{\alpha}^{\sigma}(n,E)$ being respectively
the group velocity and quantum-mechanical particle current density for the
state $\alpha$ at the energy $E$ (see Eqs. (\ref{velocity_j}%
),(\ref{QM_current})), and $V$ being the applied voltage.

\section{Results and discussion}

\subsection{Hartree approximation}

\begin{figure}[ptb]
\includegraphics[scale=0.75]{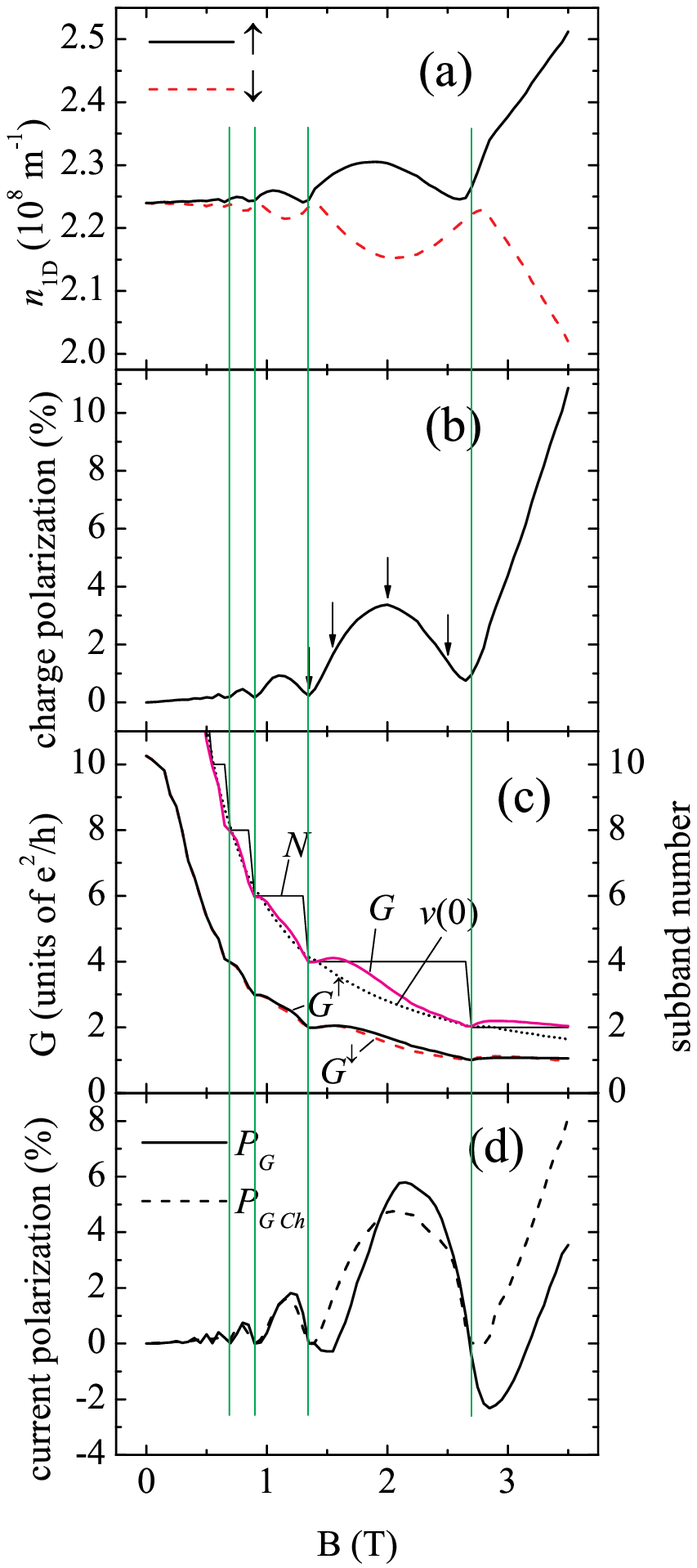}
\caption{(Color online). (a) One-dimensional charge density for
the spin-up and spin-down electrons,
$n_{1D}^{\uparrow},n_{1D}^{\downarrow}$ in the Hartree
approximation, and (b) spin polarization of the charge density,
$P_{n}=\frac{n_{1D}^{\uparrow}-n_{1D}^{\downarrow}}{n_{1D}^{\uparrow
}+n_{1D}^{\downarrow}}$, as a function of magnetic field $B$. (c)
Total number of subbands, conductance of the spin-up and spin-down
electrons, the total conductance $G=G^{\uparrow}+G^{\downarrow}$,
and the filling factor in the center of the wire $\nu(0)$ (note
$G_{\mathrm{Ch}}=\frac{e^{2}}{h}\nu(0)$. (d) The spin polarization
of $G$ and $G_{\mathrm{Ch}}$. Parameters of the wire are chosen as
indicated in the caption to Fig.~1. Arrows in (b) indicate the
magnetic fields corresponding to the magnetosubband band structure
shown in Fig. \ref{fig:Hartree_subbands}.
Temperature $T=1$K.}%
\label{fig:Hartree_polarization}%
\end{figure}

To outline the role of exchange and correlation interactions we
first study the magnetotransport in a quantum wire within the
Hartree approximation (i.e., when $V_{xc}^{\sigma}(y)$ is not
included in the effective potential (\ref{V_eff}), and the spin
polarization is driven by Zeeman splitting of the energy levels).
In our calculations we use the parameters of a quantum wire
indicated in Fig. \ref{fig:wire}, and the temperature $T=1$K. With
these parameters the effective width of the wire is $\sim 400$nm,
and the sheet electron density $n\approx 1.5 \cdot 10^{15}m^{-2}$.
Figure \ref{fig:Hartree_polarization} (a) shows the
one-dimensional (1D) electron density $n_{1D}^{\sigma}$ for the
spin-up and spin-down electrons in the quantum wire. The
pronounced feature of this dependence is a $1/B$-periodic,
loop-like pattern of the density spin
polarization $P_{n}=\frac{n_{1D}^{\uparrow}-n_{1D}^{\downarrow}}%
{n_{1D}^{\uparrow}+n_{1D}^{\downarrow}}$ as illustrated in Fig.
\ref{fig:Hartree_polarization} (b).
\begin{figure*}
\includegraphics[scale=0.8]{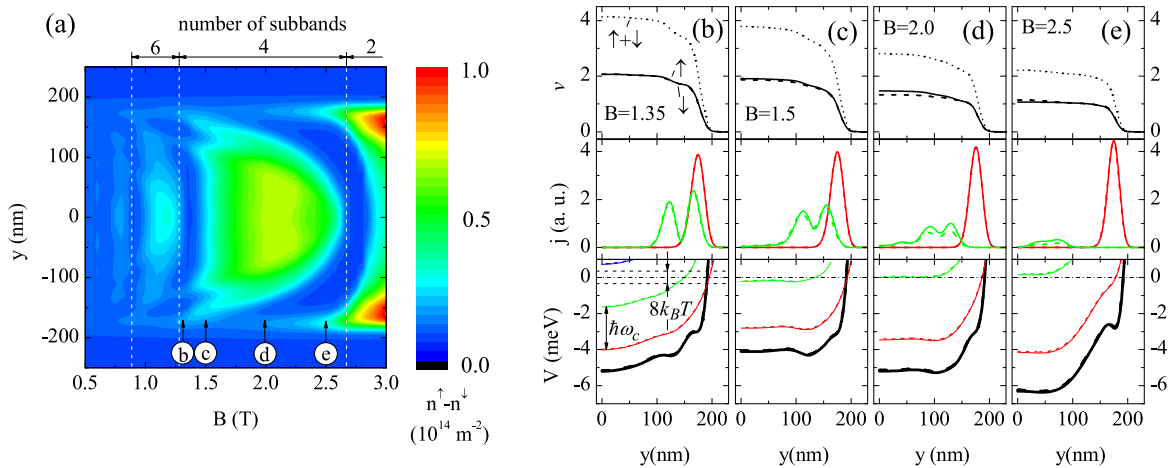}
\caption{(Color online). (a) Spatially resolved difference in the
electron densities $n^{\uparrow}(y)-n^{\downarrow}(y)$ as a
function of $B$ calculated within Hartree approximation. (b)-(e)
The subband structure for the magnetic fields indicated in (a)
(see also Fig. \ref{fig:Hartree_polarization} (b)). Upper panel:
The filling factor $\nu(y)$ for spin-up and spin-down electrons.
Middle panel: the current density distribution for spin-up and
spin-down electrons calculated according to Eq. (\ref{J}). Lower
panel: magnetosubband structure for spin-up and spin-down
electrons (solid and dashed lines correspondingly). Fat solid and
dashed lines indicate the total confining potential, Eq.
(\ref{V_total}), for respectively spin-up and spin-down electrons.
Temperature $T=1$K.}%
\label{fig:Hartree_subbands}%
\end{figure*}

Figure \ref{fig:Hartree_polarization} (c) shows the number of
spin-resolved subbands as a function of $B.$ (As the calculations
are done for the finite temperature $T,$ for a given magnetic
field, we count the subbands that lie in the energy interval
$E\lesssim E_{F}+4kT,$ where $4kT$ determines the energy window
beyond which the Fermi-Dirac distribution rapidly decays to zero).
The pronounced feature of this dependence is that the number of
subbands is always even, $N=2,$ 4, 6, \ldots\ , such that the
spin-up and spin-down subbands depopulate simultaneously. The
comparison of Figs. \ref{fig:Hartree_polarization} (a)-(c)
demonstrates that the spin polarization is directly related to the
magnetosubband structure: The polarization drops almost to zero at
the magnetic fields when the subbands depopulate. In order to
understand the origin of the spin polarization let us analyze the
evolution of the subband structure as the applied magnetic field
varies. Let us concentrate at the polarization loops in the field
interval $1.3T\lesssim B\lesssim2.6T$ when the number of the
spin-resolved subbands $N=4$ and the filling factor in the middle
of the wire $2\leq \nu (0) \leq 4$.

Figure \ref{fig:Hartree_subbands} (a) shows the spatially resolved
difference in the electron densities
$n^{\uparrow}(y)-n^{\downarrow}(y)$ as a function of $B.$ When the
subband number $N\geq4$ (for $B\lesssim2.6T$), the electron
density is mostly polarized in the inner region of the wire. We
thus concentrate first on the formation of the compressible and
incompressible strips in the inner region due to the upper
subband. Figure \ref{fig:Hartree_subbands} (b) shows the filling
factor $\nu^{\sigma}(y),$ current densities
$J_{\alpha}^{\sigma}(n),$ and the magnetosubband structure for the
magnetic field $B=1.35T.$ This field corresponds to the case when
the 5th and 6th spin-resolved subbands just became depopulated,
i.e. their bottoms are situated at $\gtrsim E_{F}+4kT$ . The 3rd
and 4th subbands are separated from the 5th and 6th by the
distance $\hbar\omega_{c}$ (with $\omega_{c}$ being the cyclotron
frequency, $\hbar\omega_{c}\gg kT$), see Fig.
\ref{fig:Hartree_subbands} (b). They are therefore situated below
the Fermy energy and are fully populated. As the results, the
electron density is constant, which corresponds to the formation
of the incompressible strip. Because of the spin-up and down
subbands are fully filled, the corresponding electron densities
are equal and the spin polarization of the electron density is
zero.

When magnetic field is raised the subbands are pushed up in the
energy and the two highest spin-resolved subbands, following
Chklovslii \textit{et al.} scenario\cite{Chklovskii}, become
pinned at the Fermi energy. The subband bottoms flatten which
signals the formation of the compressible strip in the middle of
the wire, see Fig. \ref{fig:Hartree_subbands} (c). When the
subband bottoms reach the energy $E\approx E_{F}-4kT$ , the
subbands become partially occupied. Partial subband occupation
combined with their energy separation due to Zeeman interactions
results in the different population for spin-up and down
electrons. With increase of the magnetic field the filling factor
decreases, but spin polarization increases until the subband
bottoms approach $\sim E_{F}$, Fig. \ref{fig:Hartree_subbands}
(d). This magnetic field corresponds to the maximal spin
polarization $P_n \sim 3\%$. With further increase of the magnetic
field, the subbands bottoms are pushed up above $E_{F},$ which
causes further decrease of the filling factor and diminishing
screening efficiency. As the result, the width of the compressible
strip decreases until the upper subbands become completely
depopulated and the incompressible strip forms again in the middle
of the wire, see Fig. \ref{fig:Hartree_subbands} (e). This is
accompanied by a gradual decrease of the density polarization
$P_{n}$ to zero. The shrinkage of the compressible strip in the
middle of the wire can be also clearly traced in the evolution of
the current density distribution, shown in the middle panels of
Figs. \ref{fig:Hartree_subbands} (b)-(e). It is interesting to
note that the compressible regions are not formed for the
outermost edge states corresponding to the lowest subbands $N=1$
and 2. This is because that in the field interval under study the
extension of the wave function is larger than the width of the
compressible strip predicted by the Chklovskii \textit{et al.}
theory\cite{Chklovskii}. The onset of the formation of the
compressible strips can be seen in Fig. \ref{fig:Hartree_subbands}
(e) for $B=2.5T.$ Note that the effect of the
formation/non-formation of the compressible strips in quantum
wires was discussed in details by Suzuki and Ando for the case of
spinless electrons\cite{Ando_1998}.

The described above picture of evolution of the density polarization
qualitatively holds for all other polarization loops. We stress that in all
the loops only two upper, partially occupied spin-resolved subbands contribute
to the spin polarization, whereas remaining subbands are fully (and equally)
populated and thus do no contribute to the total spin polarization. When
magnetic field exceeds $B=2.6T,$ only two subbands survive in the quantum
wire. With further increase of magnetic field the upper (spin-up) subband
gradually depopulates and the density polarization $P_{n}$ grows linearly
until it reaches 100\% when only the spin-down subband remains in the wire.

It should be also stressed that within Hartree approximation two
outermost spin-up and spin-down edge states are not spatially
polarized (i.e. they are situated at a practically same distance
from the wire edges, see Fig. \ref{fig:Hartree_subbands}).

\begin{figure}[ptb]
\includegraphics[scale=0.7]{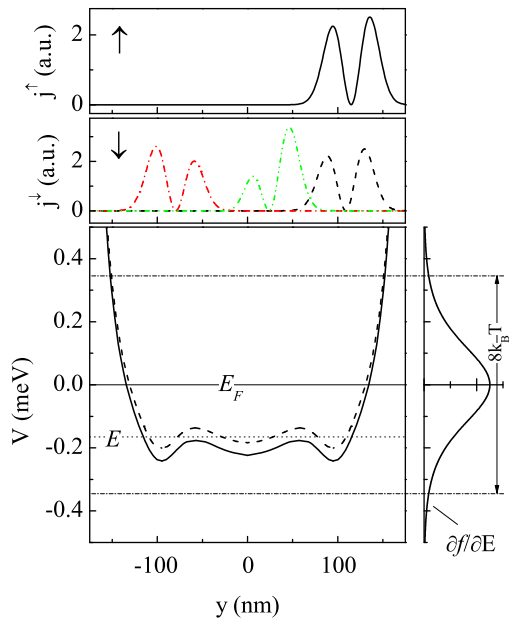}
\caption{(Color online) Lower panel: A closeup of the spin-up and
spin-down subbands $N=3,4$ (solid and dashed lines respectively)
for the magnetic field $B=1.5T$, when the polarization of the
conductance $P_{G}$ is negative. Upper panels: current density
distribution for the spin-up and spin-down electrons at the energy
$E$ indicated in the lower panel. At this energy, there is one
propagating state for spin-up electrons and three propagating
states for spin-down electrons. Left inset shows the derivative of
the Fermi-Dirac distribution determining the weight of the
contributions from the current-carrying states to the total
current density at
the given energy, see Eq. (\ref{J}).}%
\label{fig:Hartree_negative_polarization}%
\end{figure}
Figures $\ref{fig:Hartree_polarization}$ (c),(d) show the
conductance $G^{\sigma}$ for spin-up and spin-down states and its
relative spin polarization
$P_{G}=\frac{G^{\uparrow}-G^{\downarrow}}{G^{\uparrow
}+G^{\downarrow}}.$ The spin polarization $P_{G}$ follows a
similar behavior as the density polarization $P_{n}$ with one
subtle difference. Namely, the density polarization $P_{n}$ is
always positive because spin-up states always lie in energy below
the corresponding spin-down states, and, therefore
$n^{\uparrow}(y)-n^{\downarrow}(y)>0.$ In contrast, the spin
polarization of the current, after reaching zero, does not
immediately raises as the magnetic field increases, but, instead,
becomes negative before raising again. Note that this is
accompanied by a small (but noticeable) increase of the total
current (at $B\sim1.5T,$ $3T,$ see Figs.
\ref{fig:Hartree_polarization} (c)). This effect can be traced
back to the self-consistent band structure as explained below.
Figure \ref{fig:Hartree_negative_polarization} shows a closeup of
the upper subbands $N=3,4$ for the magnetic field $B=1.5T$, i.e.
when the current polarization is negative. Because the
spin-up/down subbands are not flat, for certain energies $E<E_{F}$
the upper (spin-down) subband can give rise to several propagating
states, whereas the lower (spin-up) subband corresponds to only
one propagating state, see Fig.
\ref{fig:Hartree_negative_polarization}. According to the
Landaulet formula (\ref{I}) all propagating states contribute
equally to the total current. Because of this and due to the fact
that the spin-down subband is situated closer to the Fermi energy,
the total current for the spin-down electrons is larger than the
current for the spin-up ones. This explains the negative spin
polarization of the current and the increase of the total current
at the magnetic fields just above the subband depopulation. We are
not aware of the discussion of this effect in the current
literature. The available experimental data, see e.g., Fig. 2 of
Ref. \onlinecite{Wrobel} showing a nonmonotonic dependence of the
conductance of a quantum wire as a function of magnetic field, are
consistent with the predicted behavior of the total current. Note
that this feature in the conductance also survives within the DFT
approach (see below, Fig. \ref{fig:DFT}).

Figures \ref{fig:Hartree_polarization} (c) and (d) also show the
conductance and its spin polarization calculated according to the
Chklovskii \textit{et al.} prescription \cite{Chklovskii},
$G_{\mathrm{Ch}}=\frac{e^{2}}{h}\nu(0),$ with $\nu(0)$ being the
filling factor in the center of the wire. $G_{\mathrm{Ch}}$
follows the exact conductance rather good, but does not recover
the steps in the conductance related to the subband depopulation
(see Ref. [\cite{Gerhardts_1994}] for a related discussion).
$G_{\mathrm{Ch}}$ does not also reproduce the increase of the
current and the negative conductance polarization discussed above
because these features are related to the quantum-mechanical band
structure.

As we mentioned before, the Hartree approximation predicts that spin-up and
spin-down subbands depopulate simultaneously and thus the conductance drops in
steps of $2e^{2}/h$ as the magnetic field increases. This is in strong
disagreement with the experimental observations that demonstrate that the
subbands depopulate one by one such that the conductance decreases in steps of
$e^{2}/h.$ We will show in the next section that accounting for the exchange
and correlation interactions leads to qualitatively new features in the
subband structure and brings the theory to a close agreement with experiment.

\subsection{Density functional theory in the local spin density approximation}

\begin{figure}[ptb]
\includegraphics[scale=0.7]{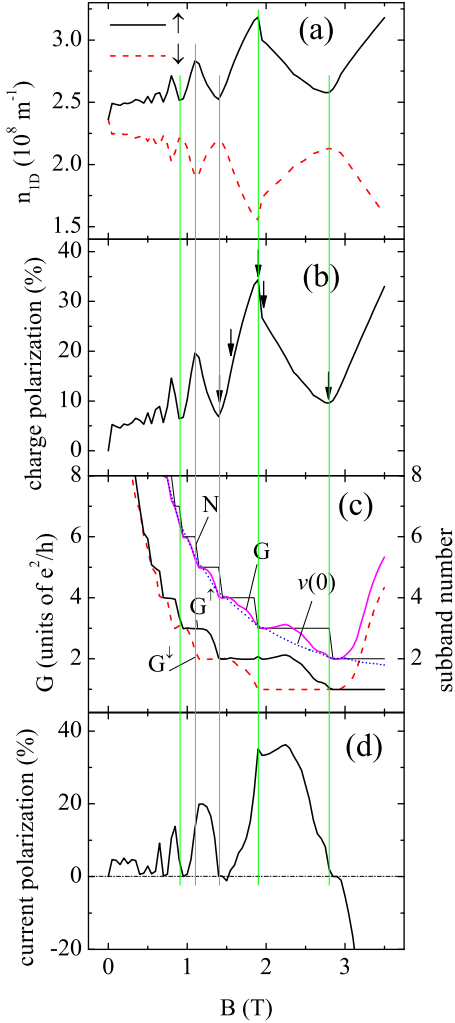}
\caption{(Color online). (a) One-dimensional charge density for
the spin-up and spin-down electrons,
$n_{1D}^{\uparrow},n_{1D}^{\downarrow}$ calculated within
DFT+LSDA, and (b) spin polarization of the charge density,
$P_{n}=\frac{n_{1D}^{\uparrow}-n_{1D}^{\downarrow}}{n_{1D}^{\uparrow
}+n_{1D}^{\downarrow}}$, as a function of magnetic field $B$. (c)
Total number of subbands, conductance of the spin-up and spin-down
electrons, the total conductance $G=G^{\uparrow}+G^{\downarrow}$,
and the filling factor in the center of the wire. (d) The spin
polarization of the conductance $P_G$. Parameters of the wire are
chosen as indicated in the caption to Fig.~1. Arrows in (b)
indicate the magnetic fields corresponding to the magnetosubband
band structure shown in Fig. \ref{fig:DFT_subbands}.
Temperature $T=1$K.}%
\label{fig:DFT}%
\end{figure}
\begin{figure*}
\includegraphics[scale=0.95]{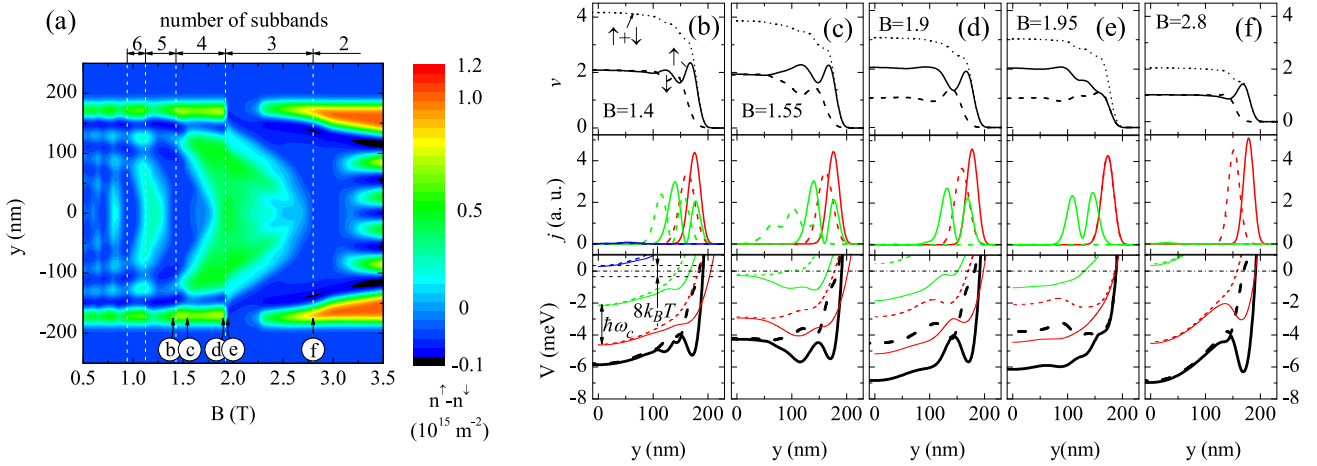}
\caption{(Color online).  (a) Spatially resolved difference in the
electron densities $n^{\uparrow}(y)-n^{\downarrow}(y)$ as a
function of $B$ calculated within DFT+LSDA. (b)-(e) The subband
structure for the magnetic fields indicated in (a) (see also Fig.
\ref{fig:DFT} (b)). Upper panel: The filling factor $\nu(y)$ for
spin-up and spin-down electrons. Middle panel: the current density
distribution for spin-up and spin-down electrons calculated
according to Eq. (\ref{J}). Lower panel: magnetosubband structure
for spin-up and spin-down electrons (solid and dashed lines
correspondingly). Fat solid and dashed lines indicate the total
confining potential, Eq. (\ref{V_total}), for respectively spin-up
and spin-down electrons. Temperature $T=1$K.}
\label{fig:DFT_subbands}%
\end{figure*}

Figures \ref{fig:DFT},\ref{fig:DFT_subbands} show the electron
density, conductance and subband structure for the quantum wire
calculated using DFT within LSDA, Eqs.
(\ref{Hamiltonian})-(\ref{LDA}). Utilization of the DFT+LSDA leads
to several major quantitative and qualitative differences in
comparison to the Hartree approximation. First, the spin
polarization of the electron density also shows a pronounced
$1/B$-periodic loop-like pattern. However, for the given magnetic
field $B$ the spin polarization in the quantum wire calculated on
the basis of DFT is of the order of magnitude higher in comparison
to the Hartree approximation. Second, the magnetosubbands
depopulate one by one, and the conductance decreases in steps of
$e^{2}/h$ (not in steps of $2e^{2}/h$ as in the case of Hartree
approach when the spin-up and spin-down subbands depopulate
simultaneously). Third, the outermost edge states become spatially
polarized (separated), which is in the strong contrast with the
Hartree approximation, where they are situated practically at the
same distance from the wire boundary.

In order to understand the effect of the exchange and correlation
interactions on the evolution of the magnetosubband structure, let
us now concentrate on the same field interval as studied in the
previous section, i.e. when the number of the subbands lies in the
interval $3\leq N\leq4$ and the filling factor in the middle of
the wire $2 \leq \nu (0) \leq 4$, see Fig. \ref{fig:DFT}. We start
from the magnetic field $B=1.4$T, where the spin polarization of
the density is minimal. Similarly to the case of Hartree
approximation (Fig. \ref{fig:Hartree_subbands} (b)), this
corresponds to the case when 5th subbands just became depopulated
as shown in Fig. \ref{fig:DFT_subbands} (b). However, in contrast
to the Hartree approximation, where the Zeeman interaction is not
strong enough to cause any significant spin polarization, in the
present case the exchange interaction leads to a
non-negligible spin polarization near the boundaries of the wire ($P_{n}%
\sim7\%$). For this magnetic field the number of subbands is even,
and the spin-up and spin-dow subbands are fully filled in the
center of the wire. As the result, the electron densities are
constant, which corresponds to a formation of the incompressible
strip in the center of the wire. Because the spin-up and spin-down
subbands are equally occupied, spin polarization in the center of
the wire is zero (Fig. \ref{fig:DFT_subbands} (a)).

\begin{figure}[ptb]
\includegraphics[scale=0.9]{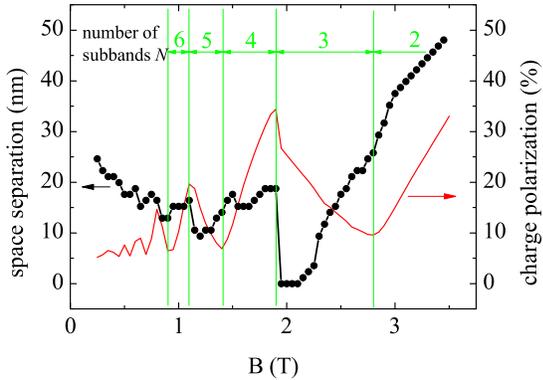}
\caption{(Color online). Spatial separation between the outermost
spin-up and spin-down edge states as a function of magnetic field
$B$. (The separation between the edge states is extracted from the
corresponding current distribution, see Fig.
\ref{fig:DFT_subbands}, middle panel).
The number of subbands and the electron density spin polarization $P_n$
from Fig. \ref{fig:DFT} is shown for comparison.}%
\label{fig:DFT_spatial_separation}%
\end{figure}

\begin{figure*}
\includegraphics[scale=0.8]{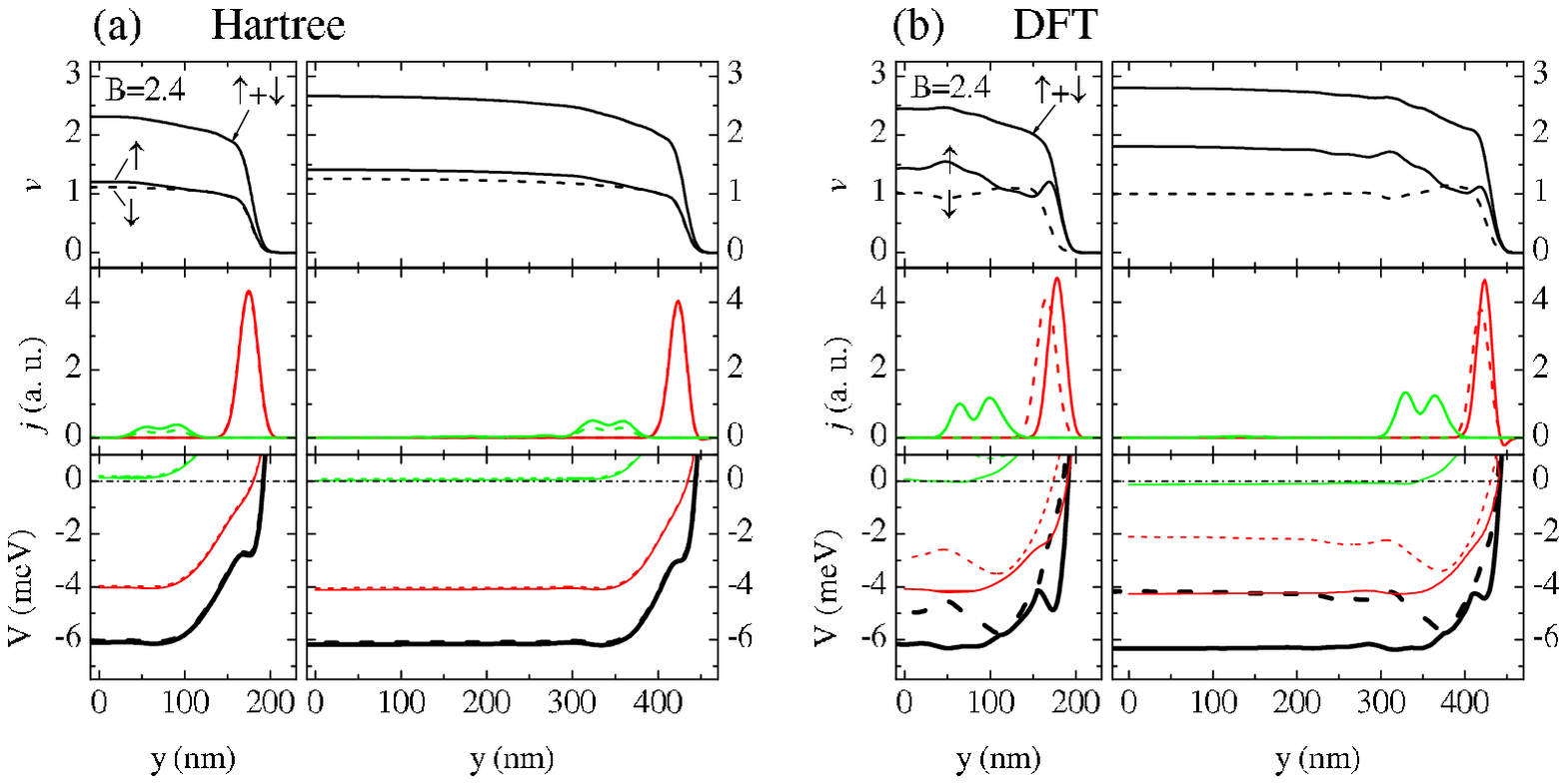}
\caption{(Color online). The filling factor, current densities and
the magnetosubband structure (upper, middle and lower panels
correspondingly) calculated within Hartree and DFT approximations
for two quantum wires with different distances between the gates,
(a) $a=500$ nm, and (b) $a=1\mu m$. Remaining parameters of the
wire are chosen as indicated in the caption to Fig.~1. Solid and
dashed lines correspond to the spin-up and spin-down states. Fat
solid and dashed lines indicate the total confining potential, Eq.
(\ref{V_total}), for respectively spin-up
and spin-down electrons. Temperature $T=1$K.}%
\label{fig:Hartree_vs_DFT}%
\end{figure*}

When the magnetic field increases, all the subbands are pushed up
in energy, and 4th subband gets pinned at $E_{F}$ near the
boundary of the wire, forming the compressible strip, Fig.
\ref{fig:DFT_subbands} (c). With increase of magnetic field, the
compressible strip extends to the center of the wire, compare
Figs. \ref{fig:DFT_subbands} (b) and (c). Note that in the Hartree
case the separation between the subbands caused by the Zeeman
splitting is small ($\ll kT$) and hence both the subbands are
pinned at $E_{F}$ (see Fig. \ref{fig:Hartree_subbands} (c)-(e)).
In contrast, in the present case only one of the subbands is
pinned at $E_{F}$ because the subband separation is determined by
the exchange interaction whose magnitude can be comparable to
$\hbar\omega_c.$

Figure \ref{fig:DFT_subbands} (d) shows the subband structure for the magnetic
field $B=1.9$T when the spin polarization of the electron density is maximal.
In this case 4th subband is about to be depopulated and all the remaining
subbands (two spin-up and one spin down) lie below $E_{F}.$ They are therefore
fully populated (2$n_{1D}^{\downarrow}\approx n_{1D}^{\uparrow}$), which
corresponds to the calculated polarization $P_{n}\sim33\%$.

When magnetic field is increased by only 0.05T, the density spin
polarization drops by $\sim10\%$, and the subband structure
experiences dramatic changes, see Fig. \ref{fig:DFT_subbands} (e).
In particularly, the spatial separation between the outermost
spin-up and spin-down states collapses from $\sim20$ nm to zero,
as shown in Fig. \ref{fig:DFT_spatial_separation}. The explanation
of this remarkable effect is based on the fact that the
electrostatic energy is dominant for the system at
hand\cite{Chklovskii}. This is illustrated in Fig.
\ref{fig:Hartree_vs_DFT} which compares the electron densities and
the magnetosubband structure in a quantum wire calculated within
the Hartree and DFT approximations for some representative value
of the magnetic field. As expected, the total electron density is
practically the same in the both approximations. At the same time,
the magnetosubbands and the spin-up and spin-down densities vary
significantly between them. It is also interesting to note that
the magnetic fields corresponding to the depopulation of even
subbands $N=2,4,6\ldots$ are practically the same with and without
accounting for the exchange and interaction terms, compare Figs.
\ref{fig:Hartree_polarization} (c) and \ref{fig:DFT} (c). The
dramatic changes in the subband structure at $B\sim1.95$ T can be
explained as follows. At $B\sim1.9$T the electron density near the
edge of the wire is dominated by spin-up electrons, see
\ref{fig:DFT} (d), the upper and middle panels. When magnetic
field is raised, 4th subband practically depopulates, and 3rd
(spin-up) subbands is pushed up in energy. As the result, the
density of the spin-up electrons associated with this subband is
redistributed towards the center of the wire. However, this small
change in the magnetic field can not affect the total density.
Because of this, the density of the remaining electrons has to be
adjusted to keep the total density unchanged. This can be done
only if the spin-down electrons associated with the subband $N=2$
are redistributed towards the edge of the wire. As a consequence
of this redistribution, the densities of the spin-up (1st subband)
and spin-down (2nd subband) electrons near the wire edge become
approximately equal and so does the total confining potential
$V^{\sigma}(n)$, Eq. (\ref{V_total}). The latter results in the
absence of the spatial separation for the outermost edge states
$N=1$ and $N=2.$ Note that the effect of a collapse of the spatial
separation between the outermost edge states is related to the
features of the quantum-mechanical band structure, and hence is
absent in the Thomas-Fermi approximation\cite{Dempsey,Stoof}. This
effect can be utilized in spintronics devices operating in the
edge state regime for injection of different spin
species\cite{Andy}.

The outermost spin-up and spin-down edge states remains spatially
degenerate up to the magnetic field $B\sim2.25$ T, see Fig
\ref{fig:DFT_subbands} (a) and Fig.
\ref{fig:DFT_spatial_separation}. The spin polarization of the
electron density $P_{n}$ gradually decreases in the range
$1.9$T$<B<2.8$T. This decrease is related to the gradual
depopulation of the 3rd (spin-up) subband. At $B\sim2.8$T this
subband practically depopulates, $P_{n}$ reaches its minimum, and
the incompressible strip is again formed in the middle of the wire
(Fig. \ref{fig:DFT_subbands} (f)). With further increase of the
magnetic field, 2nd (spin-up) subband gets pinned to $E_{F},$ and
$P_{n}$ gradually increases until it reaches 100\% at the magnetic
field when the spin-up subband depopulates.

As in the case of Hartree approximation, the evolution of the magnetosubband
structure within DFT described above qualitatively holds for all other
polarization loops.

Figure \ref{fig:DFT}(c) shows the conductance for spin-up and
spin-down electrons $G^{\uparrow},G^{\downarrow}$, the total
conductance $G=G^{\uparrow }+G^{\downarrow},$ the filling factor
in the middle of the wire $\nu(0),$ and the spin polarization of
the conductance. The total conductance $G(B)$ decreases in steps
of $e^{2}/h$ closely following the depopulation of the
magnetosubbands as $B$ increases. Note that the magnitude of
$G(B)$ in plateau regions when $N>2$ shows slight increase in
comparison to the corresponding value of $Ne^{2}/h$. This effect
has the same origin as in the case of Hartree approximation (see
Fig. \ref{fig:Hartree_negative_polarization} and a related
discussion in the text). For $N\leq2$ this effect becomes much
more pronounced in comparison to the Hartree approximation. This
is because for magnetic fields corresponding to $N\leq2,$ the
separation between bottoms of spin-up and spin-down subbands due
to the exchange interaction exceeds $8kT.$ Because the subbands
are not flat, the spin-down subband (which is pinned to $E_{F})$
gives rise to several states propagating in the bulk of the wire
as discussed in the previous section, whereas the spin-up subband
(whose bottom lies well below $E_{F})$ corresponds to only one
propagating state situated near the wire edge.

Note that the propagating states giving rise to the conductance
for $N\leq2$ are the Bloch states of an infinite quantum wire. In
a typical experimental condition, a long quantum wire is connected
to a much wider region of 2DEG\cite{Wrobel}. The edge states in
the region of 2DEG are coupled only to the edge states in the
wire. As the results, the measured conductance for $N\leq2$ does
not exhibits the increase over the plateau values of $Ne^{2}/h$
\cite{Wrobel}.

Finally, we note that our analysis of the spin polarization and
evolution of magnetosubbands in quantum wires was concentrated on
a representative wire with the distance between the gates $a=500$
nm and the sheet electron density $n\approx 10^{15}m^{-2}.$ We
would like to emphasize that all the results presented here
qualitatively hold for wires of arbitrary widths and electron
densities. This is illustrated in Fig. \ref{fig:Hartree_vs_DFT}
for the case of two quantum wires with different distances between
the gates, $a=500$ nm and $a=1\mu m,$ which shows practically
identical subband structure as well as electron and current
densities distributions.

\section{Conclusion}

In the present paper we provide a quantitative description of the structure of
edge states in split-gate quantum wires in the integer quantum Hall regime. We
start with a geometrical layout of the wire and calculate self-consistently
quantum-mechanical magnetosubband structure and spin-resolved edge states
where electron- and spin interactions are included within the density
functional theory in the local spin density approximation (DFT+LSDA).

We develop an effective and stable numerical method based on the Green's
function technique capable of dealing with a quantum wire of arbitrary width
in high perpendicular magnetic field. The advantage of this technique is that
it can be directly incorporated into magnetotransport calculations, because it
provides the eigenstates and wave vectors at a given energy, not at a given
wavevector (as conventional methods do). Another advantage of this technique
is that it calculates the Green's function of the wire, which can be
subsequently used in the recursive Green's function technique widely utilized
for magnetotransport calculations in lateral structures.

We use the developed method to calculate the self-consistent
subband structure and propagating states in the quantum wires in
perpendicular magnetic field. We discuss how the spin-resolved
subband structure, the current densities, the confining
potentials, as well as the spin polarization of the electron and
current densities evolve when an applied magnetic field varies. We
demonstrate that the exchange and correlation interactions
dramatically affect the magnetosubbands in quantum wires bringing
about qualitatively new features in comparison to a widely used
model of spinless electrons in Hartree approximation. These
features can be summarized as follows.

(a) The spin polarization of the electron density shows a
pronounced $1/B$-periodic loop-like pattern, whose periodicity is
related to the subband depopulation. For a given magnetic field
$B$ the spin polarization in the quantum wire calculated on the
basis of DFT+LSDA is of the order of magnitude higher in
comparison to the Hartree approximation (where the spin
polarization is driven by the Zeeman interaction only).

(b) The magnetosubbands depopulate one by one, and the conductance decreases
in steps of $e^{2}/h$ (not in steps of $2e^{2}/h$ as in the case of Hartree
approach when the spin-up and spin-down subbands depopulate practically
simultaneously).

(c) The outermost spin-up and spin-down edge states become
spatially polarized (separated), which is in the strong contrast
to the Hartree approximation, where they are situated practically
at the same distance from the wire boundary. We also find that the
spatial separation between the outermost edge states disappears in
the range of magnetic close to filling factor $\nu=3$ and then is
restored again when the magnetic field is raised.  This effect can
be utilized in the spintronics devices operating in the edge state
regime for injection of different spin species\cite{Andy}.

Recently, the structure of edge states around quantum antidots has
been the subject of a lively discussion\cite{adot}. Even though
the method developed in the present paper applies to quantum
wires, it is reasonable to expect that for sufficiently large
antidots (when the single particle level spacing $\Delta$ is
smaller than $kT)$ the present approach can also provide
information on the edge state structure around the antidots.

A direct probe of spin polarization of electrons in quantum dot
edge channels using polarized photoluminescence spectra has been
recently reported by Nomura and Aoyagi\cite{Nomura}. Their method
opens up a possibility for a direct probing of the electron
density spin polarization in quantum wires, such that the results
presented in our study (in particularly the spin polarization
shown in Fig. \ref{fig:DFT}(b) and Fig.
\ref{fig:DFT_subbands}(a)), can be directly verified in the
experiment.

\begin{acknowledgments}
S. I. acknowledges financial support from the Royal Swedish
Academy of Sciences and the Swedish Institute.
\end{acknowledgments}

\appendix

\section{Exchange and correlation potentials in the local spin density
approximation}

In this Appendix we provide explicit expressions for the exchange and
correlation potentials entering the DFT effective potential (\ref{V_eff}). The
exchange and correlation energies for 2DEG used in Eq. (\ref{LDA}) are given
by Tanatar and Ceperley (TC) \cite{TC}. The exchange energy reads
\begin{equation}
E_{ex}=-Ry^{\ast}\frac{4\sqrt{2}}{3\pi r_{s}}\left[  (1+\xi)^{\frac{3}{2}%
}+(1-\xi)^{\frac{3}{2}}\right]  , \label{E_ex}%
\end{equation}
where $r_{s}$ is the dimensionless density parameter which is defined in terms
of the effective Bohr radius $a_{0}^{\ast}$ (appropriate for a material with
the effective electron mass $m^{\ast}=m_{eff}m_{e}$, and the dielectric
constant $\varepsilon=\varepsilon_{r}\varepsilon_{0}),$%
\begin{equation}
r_{s}=\frac{a}{a_{0}^{\ast}},\quad a=\frac{1}{\sqrt{\pi\rho}},\quad
a_{0}^{\ast}=\frac{4\pi\varepsilon_{r}\varepsilon\hbar^{2}}{m_{eff}m_{e}e^{2}%
}=\frac{\varepsilon_{r}}{m_{eff}}a_{0},\; \label{r_s}%
\end{equation}
where Bohr radius $a_{0}=0.529\cdot10^{-9}\mathrm{\ }$m. The factor $Ry^{\ast
}=\frac{m_{eff}m_{e}e^{4}}{32\pi^{2}\varepsilon_{r}^{2}\varepsilon_{0}%
^{2}\hbar^{2}}=\frac{m_{eff}}{\varepsilon_{r}^{2}}Ry$ ($1Ry$ $=2.17989\cdot
10^{-18}$J) generalizes TC results for the case of an arbitrary effective
electron mass $m^{\ast}$ and relative dielectric constant $\varepsilon
_{r\text{ }}$, and\ converts TC expressions\cite{TC} into SI units. The
correlation energy for the unnpolarized case ($\xi=0$) and for the fully
polarized case ($\xi=1$) is approximated in the form \cite{TC},%
\begin{equation}
E_{cor}(\xi)=-Ry^{\ast}C_{0}\frac{1+C_{1}w}{1+C_{1}w+C_{2}w^{2}+C_{3}w^{3}},
\label{E_cor}%
\end{equation}
where $w=\sqrt{r_{s}}$, and the coefficients
$C_{0},C_{1},C_{2},C_{3}$ are tabulated below
\begin{table}[th]
\centering
\begin{tabular}
[c]{|l|l|l|}\hline
& unnpolarized case & fully polarized case\\
& ($\xi=0$) & ($\xi=1$)\\\hline
$C_{0}$ & -0.3568 & -0.0515\\\hline
$C_{1}$ & 1.13 & 340.5813\\\hline
$C_{2}$ & 0.9052 & 75.2293\\\hline
$C_{3}$ & 0.4165 & 37.0170\\\hline
\end{tabular}
\caption{Tabulated coefficients $C_{0},C_{1},C_{2},C_{3}$ (see text for
detail).}%
\label{tableI}%
\end{table}
For the case of an intermediate polarization, $0<\xi<1$, the correlation
energy can be interpolated between the nonpolarized and the fully polarized
cases following the receipt of von Barth and Hendin\cite{von_Barth,QDOverview}%
\begin{align}
E_{cor}(\xi)  &  =E_{cor}(0)+f(\xi)\left[  E_{cor}(1)-E_{cor}(0)\right]
\text{ }\label{interp_E}\\
\text{with }f(\xi)  &  =\frac{(1+\xi)^{\frac{3}{2}}+(1-\xi)^{\frac{3}{2}}%
-2}{2^{3/2}-2}.\nonumber
\end{align}
Taking the functional derivatives (\ref{LDA}) using the above expressions for
the exchange and correlation energies (\ref{E_ex}),(\ref{E_cor}) we arrive to
the following expression for the exchange potential $V_{ex\uparrow}$,
$V_{ex\downarrow}$, and for the correlation potentials used in Eq.
(\ref{V_eff}),%
\begin{align}
V_{ex\uparrow}  &  =-\frac{\sqrt{2}}{4}\frac{e^{2}}{\varepsilon_{0}%
\varepsilon_{r}\pi^{3/2}}\sqrt{\rho}\bigg\{\left[  (1+\xi)^{\frac{3}{2}%
}+(1-\xi)^{\frac{3}{2}}\right] \nonumber\\
&  +\frac{2\rho_{\downarrow}}{\rho}\left[  \sqrt{1+\xi}-\sqrt{1-\xi}\right]
\bigg\},\nonumber\\
V_{ex\downarrow}  &  =-\frac{\sqrt{2}}{4}\frac{e^{2}}{\varepsilon
_{0}\varepsilon_{r}\pi^{3/2}}\sqrt{\rho}\bigg\{\left[  (1+\xi)^{\frac{3}{2}%
}+(1-\xi)^{\frac{3}{2}}\right] \nonumber\\
&  -\frac{2\rho_{\uparrow}}{\rho}\left[  \sqrt{1+\xi}-\sqrt{1-\xi}\right]
\bigg\}, \label{App_V_ex}%
\end{align}%
\begin{align}
V_{cor}(\xi)  &  =V_{cor}(0)+f(\xi)\left[  V_{cor}(1)-V_{cor}(0)\right]
,\label{App_V_cor}\\
V_{cor}(\xi &  =0\text{ or }\xi=1)=\nonumber\\
&  -\frac{m_{eff}}{\varepsilon_{r}^{2}}Ry\,C_{0}\frac{1+d_{1}w+d_{2}%
w^{2}+d_{3}w^{3}+d_{4}w^{4}}{\left(  1+C_{1}w+C_{2}w^{2}+C_{3}w^{3}\right)
^{2}},\nonumber
\end{align}
where $f(\xi)$ is given by Eq. (\ref{interp_E}), and $d_{1}=2C_{1}$,$\quad
d_{2}=\left(  \frac{3}{2}C_{2}+C_{1}^{2}\right)  $,$\quad d_{3}=\left(
\frac{7}{4}C_{3}+\frac{5}{4}C_{1}C_{2}\right)  $,$\quad d_{4}=\frac{3}{2}%
C_{1}C_{3}.$

\end{document}